\documentclass[11pt,a4paper,twoside,groupcitations]{article}
\usepackage[T1]{fontenc}
\usepackage[ansinew]{inputenc}
\usepackage[english]{babel}
\usepackage{amsfonts}
\usepackage{amsmath}
\usepackage{bm}
\usepackage{array}
\usepackage{amsthm}
\usepackage{amssymb}
\usepackage{graphicx}
\usepackage{subfigure}
\usepackage{braket}
\usepackage{eucal}
\usepackage{verbatim}
\usepackage[table]{xcolor}
\usepackage{caption}
\usepackage{cite}
\usepackage{textcomp}
\usepackage{multicol}
\usepackage{tikz}
\usetikzlibrary{positioning,arrows}
\usetikzlibrary{decorations.pathmorphing}
\usetikzlibrary{decorations.markings}
\usetikzlibrary{calc,decorations.markings}
\usetikzlibrary{arrows,shapes}
\usetikzlibrary{matrix,arrows}
\usepackage{xparse}
\definecolor{jade}{HTML}{00A86B}
\newcommand{\be}{\begin{eqnarray}}
\newcommand{\ee}{\end{eqnarray}}

\renewcommand{\d}{\mbox{${\rm d}$}} 

\newcommand{\gn}{G_{\rm N}}
\newcommand{\rh}{r_{\rm H}}

\def\beq{\begin{equation}}
\def\eeq{\end{equation}}
\def\@fmsl@sh#1#2#3{\m@th\ooalign{$\hfil#1\mkern#2/\hfil$\crcr$#1#3$}}
 \def\eq#1\en{\begin{equation}#1\end{equation}}
\def\s[#1,#2]{[#1\stackrel{\star}{,}#2]}
\def\sx[#1,#2]{[#1\stackrel{\star_{x}}{,}#2]}

\def\beq{\begin{equation}}
\def\eeq{\end{equation}}

\title{\bf Binary mergers in bootstrapped Newtonian gravity: mass gap and black hole area law}
\author{Roberto~Casadio$^{ab}$\thanks{E-mail: casadio@bo.infn.it},
$\ $
Iber\^e Kuntz$^{c}$\thanks{E-mail: kuntz@fisica.ufpr.br},
$\ $
and
Octavian Micu$^{d}$\thanks{E-mail: octavian.micu@spacescience.ro}
\\
\\
$^a${\em Dipartimento di Fisica e Astronomia, Universit\`a di Bologna}
\\
{\em via Irnerio~46, 40126 Bologna, Italy}
\\
\\
$^b${\em I.N.F.N., Sezione di Bologna, I.S.~FLAG}
\\
{\em viale B.~Pichat~6/2, 40127 Bologna, Italy}
\\
\\
$^c${\em Departamento de F\'isica, Universidade Federal do Paran\'a}
\\
{\em PO Box 19044, Curitiba -- PR, 81531-980, Brazil}
\\
\\
$^d${\em Institute of Space Science}
\\
{\em P.O. Box MG-23, RO-077125 Bucharest-Magurele, Romania}
}
\begin{document}
\maketitle
\begin{abstract}
We study binary mergers in bootstrapped Newtonian gravity, where higher-order couplings are added to
the non-relativistic Lagrangian for the Newtonian potential.
In this theory, the Arnowitt-Deser-Misner (ADM) mass differs from both the proper mass of Newtonian gravity
and the proper mass of general relativity, which affects the interpretation of astrophysical and cosmological events.
The aforementioned difference particularly provides important phenomenological constraints for the mass
of the emitted matter and the compactness of the final object after the merger.
The interpretation of the GW150914 signal in this theory also shows that LIGO's findings do not violate
the mass gap, contrary to usual claims.
We indeed find that typical stellar black hole masses can fit LIGO's data for a considerable range
of compactness values.
We calculate the black hole entropy in this context, which leads to a generalised black hole area law.
Non-linear effects are found to effectively change only the gravitational strength via the renormalization
of Newton's constant in this case.
\end{abstract}  
%
%
%
%
%
%
%
\section{Introduction}
\label{sec:intro}
\setcounter{equation}{0}
One of the greatest discoveries of the past few years was the detection of gravitational waves by LIGO.
Not only has it been the strongest evidence for the gravitational radiation, but it also opened up new avenues
for astrophysical exploration.
The comparison of LIGO's data to templates from numerical relativity has shown that the observed signal
had been produced by the coalescence of a binary system of black holes. 
Adopting non-precessing numerical relativity simulations, the mass of the remnant black hole was also inferred,
culminating in the heaviest stellar black hole ever found~\cite{LIGOScientific:2016vlm,LIGOScientific:2016aoc}.
Despite the great appraisal for the discovery of gravitational waves, some portion of the community remained
skeptical due to the unusual black hole mass found after the merger.
Indeed, such a mass would violate the upper mass gap that has been found in many
contexts~\cite{Figer:2005gr,LIGOScientific:2018jsj,Woosley:2016hmi}.
\par
LIGO's data has been interpreted using general relativity, which, despite its success in the weak-field and classical regimes, has faced many downfalls at quantum scales.
Moreover, general relativity alone cannot account for the anomalous galaxy rotation curves or even dark energy.
The quest for a theory that could improve the ultra-violet behaviour of gravity or elucidate the dark sector has led
to a plethora of alternatives to general relativity.
Interpreting LIGO's data within other models could then completely change their conclusions.
\par
In this letter, we shall study the merger of compact bodies in bootstrapped Newtonian
gravity~\cite{Casadio:2018qeh,Casadio:2019cux,Casadio:2021gdf,Casadio:2020kbc,Casadio:2019pli,Casadio:2022pme},
in which the leading post-Newtonian terms are included from the onset and assumed to be of comparable
magnitude to the Newtonian one.
Because such a theory modifies the strong-field regime of Newtonian gravity, its relativistic generalization
cannot be general relativity.
The next-to-leading (and higher) order post-Newtonian terms indeed vanish identically, while the additional
coupling constants do not take the values prescribed by general relativity.
In particular, we shall show that this model can fit the GW150914 signal within the range of typical stellar mass
black holes.
\par
A major prediction of the bootstrapped program concerns the difference between the Arnowitt-Deser-Misner-like
(ADM) mass~\cite{adm} and the proper mass of a compact object.
While only the former makes its presence felt in orbits, they are both conserved (unless the system radiates),
which leads to additional constraints.
This is an important aspect of the theory that cannot be taken lightly when interpreting astrophysical phenomena.
The main goal of this paper is to investigate the phenomenological consequences of the inequivalence
between the ADM and the proper masses for the merger of compact objects.
As a result of such an inequivalence, mergers are constrained in at least two ways: (i) ejected masses cannot
be arbitrarily small in star mergers and (ii) black holes tend to become less and less compact.
The former imposes a lower bound on the mass ejected during the merger that depends on the masses
and radii of the initial objects.
The latter, on the other hand, shows that black holes are likely to merge into heavier, but less dense,
black holes.
To show this we calculate the black hole entropy using Newtonian arguments, which yields a modified law
for the area of black holes.
Such a modification results in a simple renormalization of Newton's constant, thus effectively changing
the strength of gravity.
\par
This paper is organized as follows.
In Section~\ref{s:bng}, we review the main aspects of the bootstrapped Newtonian gravity, including its field equation.
The solution of the latter is then reviewed for a source described by a homogeneous ball with negligible
pressure in Section~\ref{ss:vacuum}, where the difference between the ADM mass and the proper mass
becomes apparent.
Such a solution is then used to model stars and black holes, whose merger is studied in Section~\ref{s:mergers}.
In Section~\ref{s:thermo}, we employ Newtonian arguments to calculate the black hole entropy, generalising
the area law for black holes in the context of the bootstrapped Newtonian gravity, and show that the GW150914 event
can be accommodated without violating the upper mass gap.
Finally, we draw our conclusions in Section~\ref{sec:conc}.
\section{Bootstrapped Newtonian gravity}
\label{s:bng}
\setcounter{equation}{0}
In the most general form, the Lagrangian for the bootstrapped Newtonian potential $V=V(r)$
for a system that is both static and spherically symmetric can be written
as~\cite{Casadio:2018qeh}~\footnote{We shall use units with $c=1$.}
\be
L[V]
=
L_{\rm N}[V]
-4\,\pi
\int_0^\infty
r^2\,\d r
\left[
q_V\,\mathcal{J}_V\,V
+
q_p\,\mathcal{J}_p\,V
+
q_\rho\, \mathcal{J}_\rho \left(\rho+3\,q_p\, p\right)
\right]
\ ,
\label{LagrV}
\ee
where  
\be
L_{\rm N}[V]
=
-4\,\pi
\int_0^\infty
r^2 \,\d r
\left[
\frac{\left(V'\right)^2}{8\,\pi\,\gn}
+\rho\,V
\right]
\ ,
\label{LagrNewt}
\ee
is the standard Newtonian Lagrangian (with $f'\equiv\d f/\d r$) which yields the Poisson equation
\be
r^{-2}\left(r^2\,V'\right)'
\equiv
\triangle V
=
4\,\pi\,\gn\,\rho
\label{EOMn}
\ee
for the Newtonian potential $V=V_{\rm N}$ sourced by the matter energy density $\rho=\rho(r)$.
The gravitational self-coupling contribution is sourced by the gravitational energy $U_{\rm N}$
per unit volume~\cite{Casadio:2018qeh,Casadio:2016zpl}
\be
\mathcal{J}_V
\simeq
\frac{\d U_{\rm N}}{\d \mathcal{V}} 
=
-\frac{\left[ V'(r) \right]^2}{2\,\pi\,\gn}
\ .
\label{JV}
\ee
The static pressure $p=p(r)$ cannot be neglected for sources with relatively large compactness~\cite{Casadio:2018qeh} 
\be
X
\equiv
\frac{\gn\, M}{R}
\ ,
\label{defX}
\ee
where $M$ is the ADM-like mass that one would measure when studying
orbits~\cite{DAddio:2021xsu} and $R$ is the radius of the source. 
A potential energy $U_p$ then needs to be added such that
\be
\mathcal{J}_p
\simeq
-\frac{\d U_p}{\d \mathcal{V}} 
=
3\,p
\ ,
\label{JP}
\ee
which effectively shifts $\rho \to \rho+3\,q_p\,p$, where $q_p$ is a positive coupling constant.
These terms, together with the next-order general relativistic term $\mathcal{J}_\rho=-2\,V^2$~\cite{Casadio:2017cdv},
result in the total Lagrangian~\eqref{LagrV}.
\par
The three (dimensionless) coupling constants $q_V$, $q_p$ and $q_\rho$ could be connected to different
specific theories of the interaction between gravity and matter (for similar considerations, see, e.g.~Ref.~\cite{carloni}).
For the sake of simplicity, in the rest of this paper we shall take $q_V=q_\rho$, so that the Euler-Lagrange equation
for the bootstrapped Newtonian potential reads
\begin{equation}
	\triangle V
	=
	4\,\pi\,\gn\left(\rho+3\,q_p\,p\right)
	+
	\frac{2\,q_V\left(V'\right)^2}
	{1-4\,q_V\,V}
	\ .
	\label{EOMV2}
\end{equation}
\par
One important aspect that is not apparent from the above derivation is that the mass $M$ in Eq.~\eqref{defX}
is not equal to the source's ``Newtonian'' proper mass~\cite{Casadio:2018qeh,Casadio:2019pli}
\be
M_0
=
4\,\pi
\int_0^R
r^2\,\d r\,\rho(r)
\ .
\label{defM0}
\ee
This follows precisely because of the non-linearity of Eq.~\eqref{EOMV2}.
This contrasts the Newtonian case, where $M=M_0$, and general relativity, in which the integral~\eqref{defM0} gives
the ADM mass $M$, including all the matter energy and (negative) gravitational energy. 
The relation between $M$ and $M_0$ thus serves as a genuine prediction of bootstrapped Newtonian gravity.
However, it depends on the details of the particular source in a very non-trivial
manner~\cite{Casadio:2019cux,Casadio:2020kbc,Casadio:2019pli}.~\footnote{For energy considerations in
the bootstrapped Newtonian framework, see in particular Ref.~\cite{Casadio:2018qeh} and
Appendix~D of Ref.~\cite{Casadio:2019cux}.}
Since we are here interested in general results for binary mergers, we shall use such a relation obtained from a simple
approximation.
\subsection{Homogeneous ball in vacuum} 
\label{ss:vacuum}
In the vacuum outside a source of mass $M$ and radius $r=R$, we have $\rho=p=0$ and Eq.~\eqref{EOMV2} 
admits the exact solution~\cite{Casadio:2018qeh,Casadio:2020mch}
\be
V_{\rm out}
=
\!\!\!\!&=&\!\!\!\!
\frac{1}{4\,q_V}
\left[
1
-
\left(
1+6\,q_V\,\frac{\gn\,M}{r}
\right)^{2/3}
\right]
\ ,
\label{V_psi_N}
\ee
which is obtained by fixing the constants of integration such that the leading order in the large $r$ expansion 
reproduces the correct Newtonian behaviour $V_{\rm N}=-\gn\,M/r$ in terms of the ADM-like mass $M$.~\footnote{We
assume asymptotic flatness.}
With this choice, the first post-Newtonian term reads $V_{\rm PN} =q_V\,\gn^2\,M^2/r^2$ without any further assumption. 
A detailed investigation of the effective metric outside static spherically symmetric bootstrapped Newtonian sources 
in terms of post-Newtonian parameters (PPN) was performed in Ref.~\cite{Casadio:2021gdf}.
These results were later used to investigate geodesics and to obtain bounds using data from the Solar System and S-star
orbits near our Galaxy center~\cite{DAddio:2021xsu}.
\par
From this potential, we obtain the harmonic horizon radius where the escape velocity equals the speed
of light, $V_{\rm out}(\rh)=-1/2$~\cite{Casadio:2019cux}, which results in 
\be
\rh
=
\frac{6\, q_V \,\gn\, M}{(1+2\,q_V)^{3/2} - 1}
\ ,
\label{hrad}
\ee
and the Newtonian $\rh=2\,\gn\,M$ is recovered for $q_V\to 0$.
For $q_V=1$, one has $\rh\simeq 1.4\,\gn\,M\ge R$ for $X\gtrsim 0.7$.
We remark that, unlike general relativity, the matter core of a black hole has finite
size $R>0$ in bootstrapped Newtonian gravity (possibly as a consequence of quantum effects~\cite{Casadio:2020ueb,Casadio:2021cbv}).
Switching to areal coordinates~\cite{Casadio:2021gdf}, the above bound on the compactness of a black hole
becomes $X\gtrsim 0.8$, corresponding to the areal horizon radius $\rh\simeq 2.4\,\gn\,M$.
In the following, we shall therefore consider that black holes are characterised by $X\gtrsim 1$, and
regular stars by $X\ll 1$, for simplicity.
\par
In order to establish an explicit relation between $M$ and $M_0$, we shall model compact objets simply 
as homogenous balls with density
\be
\rho
=
\rho_0\,\Theta(R-r)
=
\frac{3\, M_0}{4\,\pi\, R^3}\, 
\Theta(R-r)
\ ,
\label{HomDens}
\ee
where $\Theta$ is the Heaviside step function enforcing the density to vanish for $r>R$ and $M_0$
is the proper mass defined in Eq.~\eqref{defM0}. 
We shall further assume that there exists a pressure profile compatible with this density.
Accounting for the pressure properly is indeed very involved and requires numerical work
(for some more details, see Refs.~\cite{Casadio:2022pme,Casadio:2020kbc}). 
\par
Any solution of Eq.~\eqref{EOMV2} for the inner potential $V_{\rm in}=V(0\le r<R)$ needs
to match smoothly with the outer vacuum solution $V_{\rm out}$ in Eq.~\eqref{V_psi_N} across the boundary
$r=R$ of the source, that is $V_{\rm in}(R)=V_{\rm out}(R)$ and $V'_{\rm in}(R)=V'_{\rm out}(R)$.
Furthermore, we are looking for potentials generated by density profiles that are finite in the centre
and the inner potential also needs to satisfy the regularity condition $V'_{\rm in}(0)=0$.
An approximate solution can be found by Taylor expanding around $r=0$ and is given
by~\cite{Casadio:2018qeh,Casadio:2019cux}
\be
V_{\rm in}
\!\!\!\!&\simeq&\!\!\!\!
\frac{1}{4\,q_V}
\left[
1- \frac{1+2\,q_V\,X\left(4-r^2/R^2\right)}{\left(1+6\,q_V\,X\right)^{1/3}}
\right]
\ .
\label{sol}
\ee
The matching conditions across the surface then yield~\footnote{In Ref.~\cite{Casadio:2022pme}, it was shown
that the internal structure of the source does not affect the potential~\eqref{M0M_V} to second order in $r$.
The ratio between $M$ and $M_0$, however, slightly changes according to the source's equation of state.} 
\be
M_0
=
\frac{M}{(1+6\,q_V\,X)^{1/3}}
\simeq
(1-2\,q_V\,X)\,M
\ ,
\label{M0M_V}
\ee
for $X\ll 1$.
A numerical analysis confirmed that this result is a good analytical approximation for objects of small and intermediate
compactness, that is for $X\lesssim 1$.
Moreover, in Ref.~\cite{Casadio:2019cux}, it was shown that for large $X$ one has
\be
M_0
\simeq
\frac{M}{q_V^{1/3}\,X^{1/3}}
\ ,
\label{M0:largeX}
\ee
which is qualitatively the limit $X\gg 1$ of Eq.~\eqref{M0M_V}.~\footnote{It was argued in Ref.~\cite{Casadio:2020ueb}
that $X$ should not be too large for black holes either, because of quantum effects
(see also Ref.~\cite{Casadio:2021cbv}), but we will not consider that bound here.} 
We note that, after the large compactness approximation \eqref{M0M_V}, $q_V$ can no longer be taken to zero,
thus the Newtonian limit is not accessible from this regime.
The phenomenological consequences of Eqs.~\eqref{M0M_V} and~\eqref{M0:largeX} will be explored in the next sections. 
\section{Merger of compact bodies}
\label{s:mergers}
\setcounter{equation}{0}
According to Eq.~\eqref{M0M_V}, if a compact object emits an amount $\delta M_0$ of proper mass,
its ADM mass will change according to
\be
\delta M_0 
=
\frac{\delta M}
{\left(1+6\,q_V\,X\right)^{1/3}}
-
\frac{2\,q_V\,X}{\left(1+6\,q_V\,X\right)^{4/3}}
\left(\delta M-X\,\frac{\delta R}{\gn}\right)
\ .
\ee
For $X\ll 1$, upon neglecting terms of order $X^2$ and higher, one obtains
\be
\delta M
\simeq
\left(1+4\,q_V\,X\right)
\delta M_0 
>
\delta M_0 
\ ,
\ee
which one could interpret as the fact that the emission of matter energy (for example, in the form of electromagnetic radiation
or massive jets) must be accompanied by the emission of gravitational waves.
On the other hand, for $X\gtrsim 1$, the object is a black hole and we expect that no classical process will
result in the emission of any form of energy, that is $\delta M_0=\delta M=0$.
\par
Let us then consider the merger of two objects of proper masses $M^{(1)}_0$ and $M^{(2)}_0$,
resulting in the formation of a new object of mass
\be
M_0^{(f)}=M_{0}^{(1)}+M_{0}^{(2)}-\delta M_0
\ ,
\ee
where $\delta M_0\ge 0$ is the amount of proper matter energy ejected in the process.
The expression for the ADM mass of the resulting object is obtained by applying~\eqref{M0M_V} to all three objects, that is 
\be
M_{(f)} 
\simeq
\left(1+6\,q_V\,X_{(f)}\right)^{1/3}
\left[
\frac{M_{(1)}}{\left(1+6\,q_V\,X_{(1)}\right)^{1/3}}
+
\frac{M_{(2)}}{\left(1+6\,q_V\,X_{(2)}\right)^{1/3}}
-\delta M_0
\right]
\ ,
\label{MfsmallX}
\ee
where $X_{(a)}=\gn\,M_{(a)}/R_{(a)}$, with $a=1,2,f$, denotes the compactness of each object.
Since $\delta M_0\ge 0$ and energy can also be emitted in the form of gravitational waves,
we expect that the difference 
\be
\!\!\!\!
\delta M 
\!\!&\simeq&\!\!
M_{(1)} + M_{(2)} - M_{(f)}  \nonumber 
\\
\!&\simeq&\!\!
\delta M_0\left(1+6\,q_V\,X_{(f)}\right)^{1/3}
\nonumber
\\
&&
+M_{(1)}\left[1- \left(\frac{1+6\,q_V\,X_{(f)}}{1+6\,q_V\,X_{(1)}}\right)^{1/3}\right]
+
M_{(2)}\left[1- \left(\frac{1+6\,q_V\,X_{(f)}}{1+6\,q_V\,X_{(2)}}\right)^{1/3}\right]
\ge 
\delta M_0
\ ,
\label{merger1}
\ee
like for a single object.
Of course, the difference with respect to that case is that two black holes with $X_{(1)}\sim X_{(2)}\gtrsim 1$
can merge and emit a finite amount of gravitational energy $\delta M>0$ without losing proper mass ($\delta M_0=0$).
Two stars, with $X_{(1)}\sim X_{(2)}\ll 1$ can instead merge and emit $\delta M \gtrsim \delta M_0>0$,
resulting in either a star with $X_{(f)}\ll 1$ or a black hole with $X_{(f)}\gtrsim 1$.
\subsection{Stars merging into stars}
\label{ss:star+star=star}
For $X_{(1)}\sim X_{(2)} \sim X_{(f)}\ll 1$, we can approximate Eq.~\eqref{merger1} by means of Eq.~\eqref{M0M_V} 
for all of the three objects involved as
\be
\delta M
\simeq
\left(1+2\,q_V\,X_{(f)}\right)
\delta M_0
-
2\,q_V\,M_{(1)}
\left(X_{(f)}-X_{(1)}\right)
-
2\,q_V\,M_{(2)}
\left(X_{(f)}-X_{(2)}\right)
\gtrsim
\delta M_0
\ ,
\ee
which yields the bound
\be
\delta M_0
\gtrsim
\left(
1
-\frac{X_{(1)}}{X_{(f)}}
\right)
M_{(1)}
+
\left(
1
-\frac{X_{(2)}}{X_{(f)}}
\right)
M_{(2)}
\ .
\label{XXX<1}
\ee
In particular, this shows that for $X_{(f)}>X_{(1)}\sim X_{(2)}$ the right hand side is always positive,
which means that some proper mass $\delta M_0$ must be expelled when a more compact object
is formed in such a merger.
We can also rewrite Eq.~\eqref{XXX<1} as 
\be
X_{(f)}
\lesssim
\frac{X_{(1)}\,M_{(1)}+X_{(2)}\,M_{(2)}}{M_{(1)}+M_{(2)}-\delta M_0}
\ ,
\ee
which shows that the amount of emitted matter energy constrains the increase in compactness.
\subsection{Stars merging into black hole}
\label{ss:star+star=BH}
For $X_{(1)}\sim X_{(2)}\ll 1$ and $X_{(f)}\gtrsim 1$, we can approximate Eq.~\eqref{merger1} 
by means of Eq.~\eqref{M0M_V} for the initial stars and Eq.~\eqref{M0:largeX} for the final black hole as
\be
\delta M_0\,q_V^{1/3} X_{(f)}^{1/3}
\!+\!
M_{(1)} \! \left[1\!-\! \left(1-2\,q_V\,X_{(1)}\right)q_V^{1/3} X_{(f)}^{1/3}\right]
\!+\!
M_{(2)} \! \left[1\!-\! \left(1-2\,q_V\,X_{(2)}\right)q_V^{1/3} X_{(f)}^{1/3}\right]
\!
\gtrsim
\delta M_0
\!
\ ,
\ee
which yields
\be
X_{(f)}^{1/3}
\lesssim
\frac{ q_V^{-1/3}\,(M_{(1)}+M_{(2)}- \delta M_0)}
{\left(1-2\,q_V\,X_{(1)}\right) M_{(1)}
+
\left(1-2\,q_V\,X_{(2)}\right) M_{(2)}
-\delta M_0}
\ .
\ee
One then find the upper bound on the final compactness
\be
X_{(f)}
\lesssim\
\frac{1}{q_V}+6\,\frac{X_{(1)}\,M_{(1)}+X_{(2)}\,M_{(2)}}{M_{(1)}+M_{(2)}-\delta M_0}
\ ,
\ee
which is in fact larger than one. 
\subsection{Star merging with a black hole}
\label{ss:star+BH=BH}
In this case we have  the initial and final black holes with $X_{(1)}\sim X_{(f)}\gtrsim 1$, while the star has $X_{(2)}\ll 1$,
and we can write Eq.~\eqref{merger1} as
\be
\delta M_0\,q_V^{1/3}\,X_{(f)}^{1/3}
+
M_{(1)}
\left(1- \frac{X_{(f)}^{1/3}}{X_{(1)}^{1/3}}\right)
+
M_{(2)}\left[1- \left(1-2\,q_V\,X_{(2)}\right)q_V^{1/3}\,X_{(f)}^{1/3}\right]
\gtrsim
\delta M_0
\ ,
\ee
which translates into the constraint
\be
X_{(f)}^{1/3}
\lesssim
\frac{q_V^{-1/3}\,(M_{(1)}+M_{(2)}- \delta M_0)}
{M_{(1)}/\left(q_V^{1/3}\,X_{(1)}^{1/3}\right)
+
\left(1-2\,q_V\,X_{(2)}\right) M_{(2)}
-\delta M_0}
\ .
\ee
\subsection{Black holes merging into black hole}
\label{ss:BH+BH=BH}
\begin{figure}[t]
\centering
\includegraphics[width=10cm]{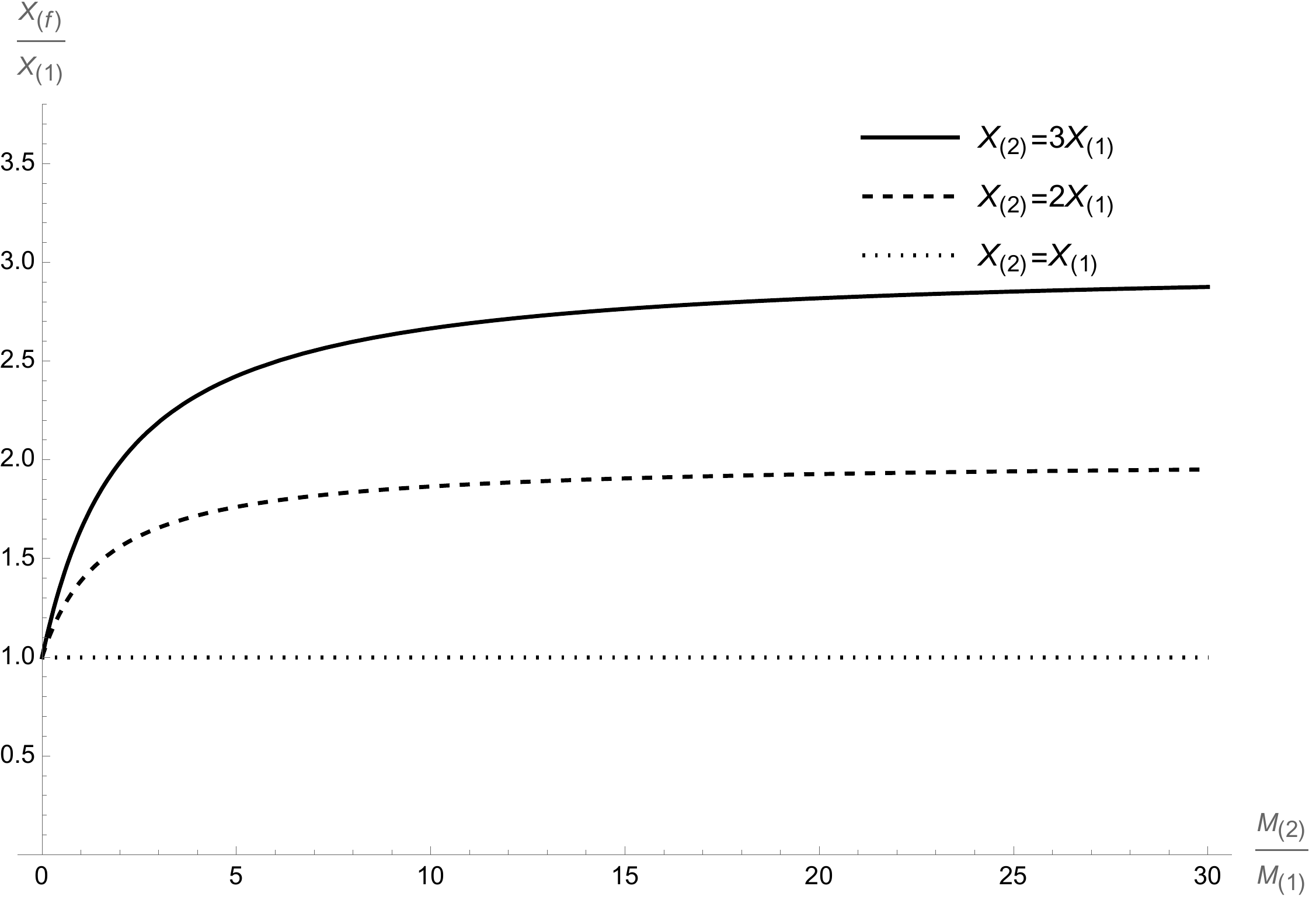}
\caption{Upper bound in Eq.~\eqref{eq:3DXf}: allowed values of
$X_{(f)}$ for given $X_{(2)}$ (both in units of $X_{(1)}\le X_{(2)}$) must lie below the corresponding curve,
which asymptotes to $X_{(2)}$ for $M_{(2)}/M_{(1)}\to\infty$.}
\label{3DXf}
\end{figure}

For $X_{(1)}\sim X_{(2)}\sim X_{(f)}\gtrsim 1$, Eq.~\eqref{M0:largeX} applies to all of the three objects involved and
we can write Eq.~\eqref{merger1} as
\be
\delta M 
\simeq
q_V^{1/3}\, X_{(f)}^{1/3}\,\delta M_0
+
M_{(1)}
\left(1- \frac{X_{(f)}^{1/3}}{X_{(1)}^{1/3}}\right)
+
M_{(2)}
\left(1- \frac{X_{(f)}^{1/3}}{X_{(2)}^{1/3}}\right)
\gtrsim
\delta M_0
\ .
\ee
Upon recalling that black holes should not emit matter energy, that is $\delta M_0=0$, we thus obtain
\be
X_{(f)}
\lesssim
\left(\frac{M_{(1)}+M_{(2)}} 
{M_{(1)}\, X_{(2)}^{1/3}+M_{(2)}\, X_{(1)}^{1/3}}
\right)^3
X_{(1)}\, X_{(2)} 
\ ,
\label{eq:3DXf}
\ee
which does not depend on $q_V$.
One can see that the above upper bound on the final compactness is always in between the two initial
values (see Fig.~\ref{3DXf}, for a few examples, in which we assumed $X_{(2)}\ge X_{(1)}$).
\par
If the two merging black holes have similar compactness $X_{(1)}\simeq X_{(2)}\equiv X_{(i)}$,
the emitted energy 
\be
\delta M 
\simeq
\left(
M_{(1)}
+
M_{(2)}
\right)
\left(
1- \frac{X_{(f)}^{1/3}}{X^{1/3}_{(i)}}
\right)
\ ,
\ee
and 
\be
X_{(f)}
\lesssim 
X_{(i)}
\ .
\ee
This means that whenever two black holes of similar compactness merge, the process will result in the formation
of a black hole which is less compact.
This upper bound will be supplemented with a lower bound in the following Section.
\section{Area law and black hole thermodynamics}
\label{s:thermo}
\setcounter{equation}{0}
Black hole mergers are the natural arena to put the laws of black hole thermodynamics to the test~\cite{Bardeen:1973gs,bekenstein}.
\par
We start from a black hole of mass $M_{(1)}\equiv M$ which absorbs a much smaller star
of mass $M_{(2)}\equiv \delta M$.
We assume that no significant amount of proper matter energy is radiated away, so that the proper mass
of the final object is the sum of the proper masses of the merging objects, 
\be
M_0^{(f)}
=
M_0^{(1)}
+
M_0^{(2)}
\ .
\ee
We will also assume that the black hole maintains approximately the same compactness, $X_{(f)}\simeq X_{(1)}\equiv X$. 
Since we have $X\gtrsim 1$ while $X_{(2)}\ll 1$, the ADM mass of the final black hole is given by
\be
M_{(f)} 
\simeq
M
+
q_V^{1/3}\,X^{1/3}
\left(1-2\,q_V\,X_{(2)}\right)
\delta M
\ .
\ee
The ``black hole area'' ${\mathcal A}=4\,\pi\,\rh^2$, with $\rh$ given in Eq.~\eqref{hrad}, under the absorption of the mass
$\delta M$ will thus change according to
\be
\frac{\Delta {\mathcal A}}{{\mathcal A}}
=
2\,\frac{M_{(f)}-M}{M}
\simeq
2\,q_V^{1/3}\,X^{1/3}\left(1- 2\,q_V\,X_{(2)}\right)\,\frac{\delta M}{M}
\ ,
\ee
which is clearly positive.
\subsection{Bootstrapped Newtonian entropy}
\label{ss:entropy}
The above result suggests that the area law holds but the precise form of the black hole entropy gets modified
in the bootstrapped Newtonian theory with respect to general relativity.
We shall then employ simple Newtonian arguments in order to study the black hole thermodynamics.~\footnote{A full account
of this topic would be obtained by quantising matter fields in the reconstructed black hole geometry~\cite{Casadio:2022pme}.}
\par
The black hole temperature is generally given by~\cite{hawking}
\begin{equation}
T = \frac{\kappa}{2\,\pi}
\ ,
\label{temp}
\end{equation}
where $\kappa = a(\rh)$ is the surface gravity, namely the gravitational acceleration on the horizon $\rh$.
In Newtonian terms, the acceleration is obtained by simply differentiating the potential $V_\text{out}$ in Eq.~\eqref{V_psi_N},
\begin{equation}
a(r)
=
\frac{\gn\, M}{r^2}
\left(
1
+ 6\, q_V \frac{\gn\, M}{r}
\right)^{-1/3}
\ .
\label{ac}
\end{equation}
From Eqs.~\eqref{hrad} and \eqref{temp} , we then find
\begin{equation}
T
=
\frac{\beta(q_V)}{8\,\pi\, \gn\, M}
\ ,
\end{equation}
where we defined
\begin{equation}
\beta(q_V)
=
\frac{
\left[
\left(1+2\,q_V\right)^{3/2}
-1
\right]^2}
{9\, q_V^2 \left(1 + 2\,q_V\right)^{1/2}}
	\ .
\end{equation}
The Hawking temperature is therefore deformed by a constant coefficient $\beta(q_V)$,
which depends only on the parameter $q_V$ and satisfies $\beta(q_V \to 0) \to 1$.
We also note that $\beta(q_V)$ effectively changes the strength of gravity, thus all corresponding thermodynamical
quantities could be obtained by simply replacing
\begin{equation}
\gn
\to
\frac{\gn}{\beta(q_V)}
\label{rule}
\end{equation}
in the standard Schwarzschild results.
\par
Indeed, the black hole entropy is obtained by considering that the absorbed heat is fully converted into the ADM mass,
\begin{equation}
\d S = \frac{\d M}{T}
\ ,
\end{equation}
thus leading to
\be
S
=
\frac{4\,\pi\, \gn\, M^2}{\beta(q_V)}
=
\beta(q_V) \frac{\mathcal A}{4\, \gn}
\ ,
\ee
which agrees with the rule~\eqref{rule}.
\par
In the following, we shall compute the difference in the entropy for the merger of two black holes assuming,
as before, that no proper matter energy is emitted (or has already been emitted before the black holes formed).
Since the entropy is additive, before the merger we have
\be
S_{(i)}
=
S_{(1)} + S_{(2)}
=
4\,\pi \left( M_{(1)}^2 + M_{(2)}^2 \right) \frac{\gn}{\beta(q_V)}
\ .
\ee
Considering the conservation of the proper mass, after the merger one finds
\be
S_{(f)}
=
\frac{4\,\pi\, \gn}{\beta(q_V)} 
\left[
\frac{X_{(f)}^{1/3}}{X_{(1)}^{1/3}}\,
M_{(1)} 
+
\frac{X_{(f)}^{1/3}}{X_{(2)}^{1/3}}\,
M_{(2)} 
\right]^2
\ .
\ee
Therefore, the variation of the entropy during the merger reads
\be
\Delta S
=
S_{(f)}
-
S_{(i)}
=
\frac{4\,\pi\, \gn}{\beta(q_V)}
\left[
\left(
\frac{X_{(f)}^{2/3}}{X_{(1)}^{2/3}}
-1
\right)
M_{(1)}^2
+
\left(
\frac{X_{(f)}^{2/3}}{X_{(2)}^{2/3}}
-
1
\right)
M_{(2)}^2 
+
\frac{X_{(f)}^{2/3}\,M_{(1)}\,M_{(2)}}
{X_{(1)}^{1/3}\,X_{(2)}^{1/3}}
\right]
\ .
\ee
Note that if the final compactness $X_{(f)}$ were larger than the compactness of each of the merging black holes
$X_{(1)}$ and $X_{(2)}$, the change in entropy would always be positive. 
However, we have seen in Section~\ref{ss:BH+BH=BH} that the compactness is supposed to decrease during the
merging and the quantity in the curly brackets above is not necessarily positive in general.
This means that a constraint on the masses and compactnesses must hold if the entropy cannot decrease.
\par
For simplicity, let us take initial black holes of approximately the same compactness $X_{(1)} \simeq X_{(2)} \equiv X_{(i)}$.
In this scenario, during a merger event, the horizon area increases if
\be
X_{(f)}^{2/3}
\left(
M_{(1)} + M_{(2)}
\right)^2 
\ge
X_{(i)}^{2/3}
\left(
M_{(1)}^2 + M_{(2)}^2
\right)
\ .
\label{constraint_S_v1}
\ee
By combining this lower bound with the upper bound from Section~\ref{ss:BH+BH=BH}, we thus find
\be
\left[
\frac{M_{(1)}^2 + M_{(2)}^2}{\left(
M_{(1)} + M_{(2)}
\right)^2 }
\right]^{3/2}
\lesssim
\frac{X_{(f)}}{X_{(i)}}
\lesssim
1
\ , 
\label{Xf:largeX}
\ee
where the lower bound is clearly less than one.
\subsection{Identical black holes mergers}
If we further take $M_{(1)}=M_{(2)}\equiv M_{(i)}$, the mass dependence from the constraint~\eqref{Xf:largeX}
goes away and the constraint becomes 
\be
\frac{1}{2^{3/2}}
\lesssim
\frac{X_{(f)}}{X_{(i)}}
\lesssim
1
\ . 
\label{Xf:largeX2}
\ee
Let us then consider a population of black holes of the same compactness $X$ and mass $M_{(i)}$.
These black holes merge in pairs repeatedly, so that at each step of the process mergers of identical black holes
with larger masses and smaller compactness take place.
After $N$ iterations, the black hole has a mass of 
\be
M_{(N)}
\simeq
2^N
M_{(i)}
\left( \frac{X_{(N)}}{X_{(i)}}\right)^{1/3}
\ .
\label{Mf:N}
\ee
The energy radiated as gravitational waves during the $N$-th merger is
\be
E_{\rm GW}
=
\delta M_N
\simeq
M_{(N)} 
-
2\,M_{(N-1)}
\simeq
2^N\,M_{(i)} 
\left( 
\frac{X_{(N)}^{1/3}}{X_{(i)}^{1/3}}
-
\frac{X_{(N-1)}^{1/3}}{X_{(i)}^{1/3}}
\right)
\ .
\ee
From the constraint~\eqref{Xf:largeX2}, we have $X_{(N)}\simeq \alpha\,X_{(N-1)}$ with $0.35\lesssim \alpha\lesssim 1$ 
and 
\be
M_{(N)}
\simeq
\left(2\,\alpha^{1/3}\right)^N
M_{(i)}
\ ,
\ee
which implies
\be
1.4^N\lesssim 
\frac{M_{(N)}}{M_{(i)}}
\lesssim 2^N
\ .
\ee
\subsection{GW150914 signal and the upper mass gap violation}
\label{ssec:mergers2}
From Eq.~\eqref{merger1}, the energy emitted into gravitational waves for large compactnesses, reads
\be
E_{\rm GW}
=
\delta M
\simeq
M_{(1)}
\left(
1
-
\frac{X_{(f)}^{1/3}}{X_{(1)}^{1/3}}
\right)
+
M_{(2)}
\left(
1
-
\frac{X_{(f)}^{1/3}}{X_{(2)}^{1/3}}
\right)
\ .
\label{gwe:largeX}
\ee
Note that, although mergers always yield heavier black holes, the final compactness must be smaller according
to Section~\ref{ss:BH+BH=BH}.
\par
The relation~\eqref{MfsmallX} for very compact objects must be computed using Eq.~\eqref{M0:largeX} and reads
\be
M_{(f)} 
\simeq
X_{(f)}^{1/3}
\left[
\frac{M_{(1)}}{X_{(1)}^{1/3}}
+
\frac{M_{(2)}}{X_{(2)}^{1/3}}
\right]
\ .
\ee
Using the masses observed at LIGO, that is $M_{(1)} = 29\, M_\odot$, $M_{(2)} = 36\, M_\odot$ and
$M_{(f)} = 62\, M_\odot$~\cite{LIGOScientific:2016aoc}, leads to
\be
62
\simeq
29 \left( \frac{X_{(f)}}{X_{(1)}}\right)^{1/3}
+
36
\left( \frac{X_{(f)}}{X_{(2)}}\right)^{1/3} \ .
\ee
Since the two initial black holes are similar in size, we approximate $X_{(1)} \sim X_{(2)}\equiv X_{(i)}$ to wit
\be
\frac{X_{(f)}}{X_{(i)}}\simeq 0.87
\ .
\ee
This ratio is in the allowed range in Eq.~\eqref{Xf:largeX2}. 
It also shows that the error induced by assuming $X_{(1)} \simeq X_{(2)}$ was fairly small since the compactness
decreases by $0.13\%$ when the black hole mass is doubled. 
\par
From Eq.~\eqref{M0:largeX}, one finds that typical stellar black hole (proper) masses fall in the range
\begin{equation}
	5\, M_\odot \lesssim \frac{M_{(i)}}{q_V^{1/3} {X_{(i)}}^{1/3}} \lesssim 50\, M_\odot \, ,
	\label{massgap}
\end{equation}
as expected from the lower and upper mass gaps \cite{Figer:2005gr,LIGOScientific:2018jsj,Woosley:2016hmi}.
The measurement of a $62\, M_\odot$ black hole by LIGO is often used as evidence for the absence of an upper mass gap.
Nevertheless, the difference between ADM and proper masses that originates from the aforementioned non-linearities
requires some caution before jumping to such strong conclusions.
The range~\eqref{massgap} can indeed be translated into a bound on the compactness $X_{(i)}$, namely
\begin{equation}
	\frac{1}{q_V}
	\left(
		\frac{M_{(i)}}{50 M_\odot}
	\right)^3
	\lesssim X_{(i)} \lesssim
	\frac{1}{q_V}
	\left(
		\frac{M_{(i)}}{5 M_\odot}
	\right)^3
	\ .
	\label{bound}
\end{equation}
The parameter $q_V$ is strongly constrained to be closed to one \cite{DAddio:2021xsu}, thus in the following we shall take $q_V \sim 1$.
Applying Eq.~\eqref{bound} for each of the masses observed at LIGO
($M_{(1)} = 29\, M_\odot$, $M_{(2)} = 36\, M_\odot$ and $M_{(f)} = 62\, M_\odot$)~\cite{LIGOScientific:2016aoc},
one finds
\begin{equation}
	1.9 \lesssim X_{(i)} \lesssim 195.1
	\ .
	\label{numbounds}
\end{equation}
We stress that LIGO findings do not violate the upper mass gap in the range \eqref{numbounds}.
Non-linear effects are thus able to fit LIGO data without requiring untypical values for stellar black hole masses. 
\section{Conclusions}
\label{sec:conc}
\setcounter{equation}{0}
In this work, we have studied binary mergers in the framework of the bootstrapped Newtonian gravity.
An interesting prediction of this theory regards the difference between the ADM and proper mass of an object.
This feature is particularly important when interpreting astrophysical and cosmological phenomena because they are both conserved, yielding additional constraints..
The final ADM black hole mass after the emission of the GW150914 measured by LIGO can thus satisfy
the proper mass gap found for typical scenarios of stellar black holes.
\par
The study of mergers has also revealed several constraints that showed up due to the ADM-proper mass
inequivalence.
When stars merge into stars, one such constraint imposes a lower bound on the ejected mass during
the merger, which depends on the inherent properties of the initial bodies.
For stars merging into black holes, the ejected mass is rather limited from above.
Similar lower/upper bounds take place when stars merge with black holes and for black hole mergers.
In each of these cases we also calculate constraints on the compactness of the final object in terms of the initial ones.
We stress that these bounds are testable predictions of the theory, which can be falsified by the observation
of an ejected mass smaller (respectively greater) than the lower (respectively upper) bounds.
\par
For the case of black holes, we used Newtonian arguments to calculate the black hole entropy.
The result turned out to take the same form as the celebrated area law, apart from an additional
multiplicative constant factor which effectively alters the gravitational strength. 
The second law of thermodynamics applied to black hole mergers, in which no mass (but only radiation)
is emitted, implies that when such processes take place, an additional constrain must apply besides
the decrease of the compactness after the merger.
The final black hole is thus heavier but less dense than the initial ones.
\section*{Acknowledgments}
R.C.~is partially supported by the INFN grant FLAG and his work has also been carried out in
the framework of activities of the National Group of Mathematical Physics (GNFM, INdAM).
O.M.~was supported by the Romanian Ministry of Research, Innovation, and Digitisation, grant no.~16N/2019
within the National Nucleus Program.
%
%
%
%

%
\end{document}